\documentclass[twocolumn,preprintnumbers,amsmath,amssymb]{revtex4}
\usepackage{amsfonts}
\usepackage{amsmath,amsfonts,amssymb}
\usepackage{bbm}
\usepackage{graphicx,epsfig}
\usepackage{bm}
\usepackage{dcolumn}

\begin{document}

\title{Quantum search with interacting Bose-Einstein condensates}

\author{Mahdi Ebrahimi Kahou and David L. Feder}
\affiliation{Institute for Quantum Information Science and Department of 
Physics and Astronomy, University of Calgary, Calgary T2N 1N4, Alberta, Canada}

\date{\today}

\begin{abstract}
One approach to the development of quantum search algorithms is the quantum 
walk. A spatial search can be effected by the continuous-time evolution of a 
single quantum particle on a graph containing a marked site. In 
many physical implementations, however, one might expect to have multiple 
particles. In interacting bosonic systems at zero temperature, the dynamics 
is well-described by a discrete nonlinear Schr\" odinger equation. We 
investigate the role of nonlinearity in determining the efficiency of the 
spatial search algorithm within the quantum walk model, for the complete 
graph. The analytical calculations reveal that the nonlinear search time scales 
with size of the search space $N$ like $\sqrt{N}$, equivalent to the linear 
case though with a different overall constant. The results indicate that 
interacting Bose-Einstein condensates at zero temperature could be natural 
systems for the implementation of the quantum search algorithm.
\end{abstract}

\maketitle
\section{\label{Intro}Introduction}
Searching is arguably one of the most important problems in computer science.
The search problem consists of finding a particular (marked) element 
in a set containing $N$ items. The simplest classical approach is to uniformly 
sample the set until the marked element is found, which on average occurs in
a time $t=O(N)$. Though the sampling can be recast into the framework of
Markov chains, the $O(N)$ scaling of the classical search time is optimal. 

Grover~\cite{Grover1,Grover} constructed an algorithm based on quantum 
mechanics that can find the marked element of a set quadratically faster, in a 
time $t=O(\sqrt{N})$. This was proven to be optimal~\cite{Bennett1997}. The 
quantum search algorithm can be recast in terms of quantum walks, the quantum 
mechanical extension to Markov chains. The origins of the discrete-time quantum 
walk can be traced to Meyer~\cite{Meyer1996a,Meyer1996b}, and the concept was 
further developed in 2001 by several 
others~\cite{Watrous2001,Ambainis2001,Aharonov2001}. The continuous-time 
quantum walk, the quantum analog of a continuous-time Markov process, was 
introduced by Farhi and Gutmann~\cite{Farhi1998} and extended by them and 
Childs~\cite{Childs2002}. Numerous algorithms based on the quantum walks were 
soon developed that were shown to be more efficient than the best-known 
classical 
algorithms~\cite{Kempe2003,Ambainis2003,Kendon2006,Reitzner2011,Venegas2012}. 
Among these are quantum walk search algorithms, based both on the 
discrete-time quantum 
walk~\cite{Shenvi2003,Szegedy2004,Ambainis2005,Santha2008,Reitzner2009} and on 
the continuous-time quantum walk~\cite{Childs2004a,Childs2004b,Agliari2010}. In 
these approaches, the set corresponds to the vertices of a graph, and the 
marked element is one distinguishable vertex. While the spatial search time 
attains the optimal scaling on most graphs, including the complete graph,
complete bipartite graphs, $m$-partite graphs~\cite{Reitzner2009}, the 
hypercube~\cite{Shenvi2003}, the Johnson graph~\cite{Magniez2011}, and regular 
lattices with dimension equal to three or greater~\cite{Childs2004b}, it 
remains at best $t=O(\sqrt{N\log N})$ for quantum walks on the two-dimensional 
square lattice despite much 
effort~\cite{Childs2004b,Ambainis2005,Tulsi2008,Lovett2012}, and $t=O(N)$ in 
one dimension.

Quantum walks have been realized using a variety of experimental approaches. 
Some of the earliest experiments were based on nuclear magnetic 
resonance~\cite{Du2003,Ryan2005}. Three steps of a discrete-time quantum walk 
were realized with $^{25}$Mg$^+$ ions in a linear Paul trap~\cite{Schmitz2009};
longer quantum walks were effected more recently with $^{40}$Ca$^+$ 
ions~\cite{Zahringer2010}. Five steps of a quantum walk were implemented using
passive optical elements~\cite{Schreiber2010,Broome2010}.
Quantum walks have also been implemented with single neutral $^{133}$Cs 
atoms~\cite{Karski2009} confined in optical lattices~\cite{Bloch2008}.

Because the behavior of quantum walks is governed by quantum interference, it 
is not necessary to restrict physical systems to single walkers. For example, a 
quantum walk with two identical photons was demonstrated using evanescently 
coupled waveguides~\cite{Perets2008,Peruzzo2010}. In fact, are several 
indications that quantum walks with multiple indistinguishable particles have 
unique properties. Non-classical correlations arise between two non-interacting 
photonic 
walkers~\cite{Peruzzo2010,Broomberg2011,Owens2011,Meinecke2013,Crespi2013}. 
Two-photon quantum walks with conditioned interactions and strong 
nonlinearities were recently reported~\cite{Schreiber2012}. Quantum walks have
also been realized using Bose-Einstein condensates of 
$^{87}$Rb~\cite{Mandel2003}. Some classically intractable problems, such as 
boson sampling, are efficiently
solved using quantum walks with multiple indistinguishable 
bosons~\cite{Broome2013,Spring2013}. Theoretical work suggests that multiple
indistinguishable walkers could help determine if two graphs are 
isomorphic, with interactions improving the power of the 
algorithm~\cite{Gamble2010,Berry2011,Rudinger2012}. 
For suitably defined graphs, quantum walks with multiple interacting walkers 
are able to perform arbitrary quantum algorithms~\cite{Childs2013}.


Large numbers of indistinguishable bosons at low temperatures can form a 
Bose-Einstein condensate (BEC). The implementation of a continuous-time 
quantum walk using BECs is equivalent to allowing the bosons to evolve under a 
governing lattice or graph Hamiltonian. In the presence of weak particle 
interactions, the resulting Gross-Pitaevskii (GP) equation of motion in the 
mean-field approximation is nonlinear~\cite{PethickSmith}. In principle, the 
presence of 
nonlinearity in quantum dynamics can radically alter the performance of quantum 
algorithms~\cite{Abrams1998}, even allowing NP-complete problems to be solved 
in polynomial time. One might therefore conjecture that the timescale for the 
quantum search problem could be modified by the presence of nonlinearity. That
said, the nonlinearity that appears in the GP equation has its 
origins in ordinary linear quantum mechanics, which would appear to rule out 
any improvement in the scaling of the quantum search time with $N$ (though it 
could always be worse). 
In any case, it is important to know how the presence of nonlinearity in the 
governing equations would affect the performance of a quantum search. The 
influence of (a physically motivated) nonlinearity on the time to effect a 
quantum walk spatial search is the central question addressed in this work. 
The results indicate that interacting BECs can indeed implement a quantum 
spatial search algorithm.

We consider the quantum search algorithm on the complete graph using a 
continuous-time quantum walk based on the discrete GP equation. 
In the complete graph, each site or vertex is connected to every other; a boson
at a given site can tunnel or hop to any other site with equal probability.
While a physical lattice with the connectivity of the complete graph has not 
yet been realized experimentally, a recent theoretical proposal to simulate the
hypercube graph with ultracold atoms~\cite{Boada2012} suggests that other 
graphs with unusual connectivity properties might be experimentally feasible in 
the future. In any case, the study of the complete graph offers several 
theoretical advantages. The search time obtained in the linear quantum walk 
algorithm has previously been shown to be optimal~\cite{Childs2004a}. The 
neighborhood of every vertex corresponds to all other vertices, so that the 
quantum walk naturally mimics a uniform sampling of the set elements. Last, the 
symmetry of the complete graph allows the vertex set to be decomposed into two 
inequivalent elements: the marked vertex, and the set of unmarked vertices. 
This allows the $N$-dimensional Hilbert space to be reduced to two dimensions, 
greatly simplifying the analysis of the nonlinear problem.

This manuscript is organized as follows. In Sec.~\ref{background} we review the
continuous-time quantum walk approach to the spatial search problem, and
derive the associated nonlinear equation of motion for a quantum search based
on interacting Bose-Einstein condensates. The analytical results are 
presented in Sec.~\ref{nlqw}, and the criteria for a complete search (unit
output probability on the marked vertex in the limit of large $N$) and an 
incomplete search are derived. The performance of the algorithm in the presence
of errors is analyzed numerically in this section. The results are summarized 
in Sec.~\ref{concl}.

\section{Background}
\label{background}

\subsection{Continuous-time quantum walk search algorithm}

In the continuous-time quantum walk, the state of the quantum walker 
$|\psi\rangle$ is evolved in time by the action of the Hamiltonian
$H_0=-\gamma L$, where $L$ is the Laplacian operator and $\gamma$ is the 
transition amplitude. Given an $N$-dimensional graph $G=(V,E)$ where 
$V=\{1,2,\ldots,N\}$ and $E$ correspond to the set of vertices and edges 
respectively, one can define the Laplacian as $L=A-D$, where $A=A(G)$ is the 
adjacency matrix associated to the graph $G$ and $D$ is a diagonal matrix whose 
elements are $D_{ii}=\sum_jA_{ij}={\rm deg}(i)$, the degree of vertex 
$i$ (the inclusion of the diagonal is not needed if the graph is regular).
The adjacency matrix specifies the graph connectivity and its matrix elements 
are defined as
\begin{equation}
A_{ij}=\begin{cases}
1 & (i,j)\in E\cr 0 & \mbox{otherwise}.\cr
\end{cases}
\end{equation}
In the continuous-time quantum walk, one associates each vertex $i$ to a basis 
vector $|i\rangle$; the
set of basis vectors spans the $N$-dimesional Hilbert space. The state of the 
quantum walker is $|\psi(t)\rangle=\sum_ia_i(t)|i\rangle$, where $a_i(t)$ are 
time-dependent complex coefficients. The quantum walk is then effected by 
performing 
the unitary transformation $|\psi(t)\rangle=e^{-\imath H_0t}|\psi(0)\rangle$ 
on the particle state vector ($\hbar$ is set to unity in this work).

In the continuous-time quantum walk search algorithm of Childs and 
Goldstone~\cite{Childs2004a}, one of
the basis vectors $|w\rangle$ is treated differently. This is accomplished by 
introducing a marking or oracle Hamiltonian:
\begin{equation}
H_w\equiv -|w\rangle\langle w|.
\end{equation}
The quantum state is then evolved according to the total Hamiltonian
$H=H_0+H_w$, i.e.\ $|\psi(t)\rangle=e^{-\imath Ht}|\psi(0)\rangle$. It was 
shown that if the initial state is chosen to be the uniform superposition of 
all sites
\begin{equation}
|\psi(0)\rangle=|S\rangle\equiv\frac{1}{\sqrt{N}}\sum\limits_{i=1}^{N}
|i\rangle,
\label{init}
\end{equation}
then there exists a time $t_s$ and value of $\gamma$ for which the probability
on the marked site $|\langle w|\psi(t_s)\rangle|^2=|\psi_w(t_s)|^2$ will attain 
unity. For the complete graph, $t_s=\frac{\pi}{2}\sqrt{N}$ and 
$\gamma=\frac{1}{N}$. For the hypercube and an $m$-dimensional square lattice 
for $m>4$, the search time remains $t_s=O(\sqrt{N})$.

\subsection{Discrete Gross-Pitaevskii equation}

A convenient starting point for the description of interacting BECs is the 
Gross-Pitaevskii equation~\cite{PethickSmith}
\begin{equation}
\imath\frac{\partial}{\partial t}\Psi({\bf r},t)
=\left[-\frac{1}{2m}\nabla^2+V({\bf r})+U|\Psi({\bf r})|^2\right]
\Psi({\bf r},t),
\label{eq:GP}
\end{equation}
where $V({\bf r})$ is some time-independent external potential, $U$ is the 
particle interaction strength, and $\Psi({\bf r},t)$ is the BEC wavefunction.
All of the $M$ bosons are assumed to be in the same single-particle state, so
the normalization condition is $\int d{\bf r}|\Psi({\bf r},t)|^2=M$. It is 
convenient to define the BEC wavefunction in terms of a new wavefunction 
$\Psi({\bf r},t)=\sqrt{M}\psi({\bf r},t)$ so that $\int d{\bf r}
|\psi({\bf r},t)|^2=1$. 

Consider functions $V({\bf r})$ corresponding to a series of $N$ potential 
energy wells, each centered at ${\bf r}_j$ with $j=1,\ldots,N$. A simple 
example in one dimension would be $V(x)=V_0\cos^2(\pi x/a)$ in a box of length
$L$, where $a$ is some arbitrary length scale and $0\leq x\leq L$; if $L/a$ is
an integer, the potential minima are found at $x/a=(2n+1)/2$ with $n$ positive 
integers and $n_{\rm max}=N= L/a$. If the confinement is sufficiently strong,
the particles comprising the BEC will be completely confined to the sites of 
the potential $V({\bf r})$; reducing the confinement would then allow tunneling 
between nearby sites. The BEC wavefunction can then be expanded in a basis of 
Wannier functions $w({\bf r}-{\bf r}_j)$ localized to the sites,
$\psi({\bf r})=\sum_j\psi_jw({\bf r}-{\bf r}_j)$. Inserting this into
Eq.~(\ref{eq:GP}), multiplying on the left by $w^*({\bf r}-{\bf r}_k)$ and 
integrating over all space gives
\begin{equation}
\imath\frac{\partial}{\partial t}\psi_k=-\sum_j\gamma_{kj}\psi_j+g_k|\psi_k|^2
\psi_k,
\label{eq:GP2}
\end{equation}
where
\begin{equation}
\gamma_{kj}=-\int d{\bf r}w^*({\bf r}-{\bf r}_k)\left[-\frac{1}{2m}
\nabla^2 +V({\bf r})\right]w({\bf r}-{\bf r}_j)
\end{equation}
is the amplitude to tunnel between sites centered at ${\bf r}_j$ and 
${\bf r}_k$, and
\begin{equation}
g_k=MU\int d{\bf r}|w({\bf r}-{\bf r}_k)|^4
\end{equation}
is the on-site interaction strength. In deriving Eq.~(\ref{eq:GP2}), the 
Wannier functions are assumed to be orthonormal, and to be so strongly 
localized that the spatial integrals of four Wannier functions are 
insignificant unless all their arguments are the same. 

For all the analytical calculations in this work, we will make the further 
simplifying assumption that $\gamma_{kj}=\gamma>0$ for all nearest neighbors 
$(j,k)\in E$ and that the particle interactions are site-independent $g_k=g$. 
For mean field theory to remain valid, one requires $g/\gamma\lesssim 5.8z$,
where $z$ is the site coordination number (vertex degree) in the limit 
$z\to\infty$~\cite{Fisher1989,Sheshadri1993}. For the complete graph $K_N$ 
with $N=|V|$ vertices investigated in the present work, each vertex has 
$z=N(N-1)/2$ neighbors. Mean-field theory therefore requires 
$g/\gamma\lesssim N(N-1)$ which is easy to satisfy for large $N$.

The discrete GP equation can then be written
\begin{equation}
\imath\frac{\partial}{\partial t}\psi_k=-\gamma A_{kj}\psi_j
+g|\psi_k|^2\psi_k,
\label{MFdyn2}
\end{equation}
where $A_{kj}$ are the matrix elements of the graph adjacency matrix $A$ 
defining the connectivity of the sites. For graphs with constant degree (i.e.\
site valency) $d$, the Laplacian is $L=A-D=A-dI$. Because a constant energy 
offset cannot change
the dynamics, the BEC wavefunction is equivalently described by the GP 
Hamiltonian
\begin{equation}
H_{\rm GP}=-\gamma L+g\sum_{k=1}^N|\psi_k|^2|k\rangle\langle k|
\label{HGP}
\end{equation}
which generates time evolution via the usual Schr\" odinger equation
$\imath\partial\psi_k/\partial t=\langle k|H_{\rm GP}|\psi\rangle$, where
$\psi_k=\langle k|\psi\rangle$.
The non-linear quantum walk search Hamiltonian then takes the following form: 
\begin{equation}
H=H_{\rm GP}-|w\rangle\langle w|=-\gamma L-|w\rangle\langle w|
+g\sum\limits_{i=1}^{N}|\psi_{i}|^2|i\rangle\langle i|.
\label{Hsearch}
\end{equation}
For example, in ultracold atom experiments an individual site of an optical 
lattice could in principle be `marked' by modifying the local potential using 
single-site addressing~\cite{Sherson2010}.

The simplest (though not the only) way to guarantee that $\gamma_{jk}=\gamma$ 
is to assume that $|{\bf r}_j-{\bf r}_k|=$~constant for all nearest neighbors 
$(j,k)\in E$. The dimension ${\rm dim}(G)$ of a graph $G$ is defined as the 
smallest number $D$ for which the graph satisfying this property can be 
embedded in $D$-dimensional Euclidean space ${\mathbb R}^D$. For the special 
case $D=2$ these graphs are called unit-distance graphs; examples include 
cycles, regular two-dimensional lattices, and hypercubes. Of course, the 
three-dimensional regular lattice also has dimension $D=3$. The complete graph
$K_N$ with $N=|V|$ vertices investigated in the present work unfortunately has 
dimension $d=N-1$~\cite{Erdos1961}; the vertices form a $D$-simplex arranged 
over the $D$-dimensional hyperspherical surface of circumradius $r$ with fixed 
distance $r\sqrt{2(N+1)/N}$~\cite{Fiedler2011}. This makes the direct physical 
realization of the complete graph connectivity via a potential energy function 
$V({\bf r})$ challenging unless a three-dimensional embedding can be found. In 
principle this might be possible by varying the positions of the physical sites 
while simultaneously changing the potential barrier heights. A much simpler 
approach would likely be to pursue a photonic implementation employing passive 
optical elements such as multiple beam splitters~\cite{Lu1989}. The complete 
graph is nevertheless interesting from a purely theoretical perspective.

\section{Nonlinear quantum walk search on the complete graph}
\label{nlqw}

\subsection{Reduction to two dimensions}

Consider the action of the Hamiltonian~(\ref{Hsearch}) for the $N$-vertex
complete graph $K_N$. The complete graph is associated to the adjacency matrix
with elements $A_{ij}=1-\delta_{ij}$. The vertex degree is constant $d=N-1$ so
the Laplacian is $L=A-(N-1)I=J-NI$, where $J$ is the $N$-dimensional all-one 
matrix. In terms of the initial state $|S\rangle$ defined in Eq.~(\ref{init}), 
the Laplacian is $L=N|S\rangle\langle S|-NI$. Since $-NI$ corresponds to a 
constant energy shift it can not change the dynamics of the system. The 
Hamiltonian for the quantum walk search~(\ref{Hsearch}) then becomes
\begin{equation}
H=-\gamma N |S\rangle\langle S| -|w\rangle\langle w|
+g\sum\limits_{i=1}^{N}|\psi_{i}|^2|i\rangle\langle i|.
\label{Hsearch2}
\end{equation}
In the absence of the nonlinear term, this Hamiltonian corresponds to a 
two-level operator for the states $|S\rangle$ and $|w\rangle$. The Hamiltonian
then rotates $|S\rangle$ into $|w\rangle$ in time $t_s\sim\sqrt{N}$ inversely
proportional to the two states' overlap $\langle S|w\rangle=1/\sqrt{N}$. 

It would is desirable to express the nonlinear Hamiltonian~(\ref{Hsearch2}) as 
a two-level operator. For the complete graph, the initial 
condition~(\ref{init}) can be written as
\begin{equation}
|\psi(0)\rangle=|S\rangle=\frac{1}{\sqrt{N}}\left(|w\rangle+\sqrt{N-1}
|\alpha\rangle\right),
\end{equation}
where 
\begin{equation}
|\alpha\rangle\equiv\frac{1}{\sqrt{N-1}}{\sum_i}'|i\rangle
\end{equation}
labels the state orthogonal to $|w\rangle$ corresponding to the superposition
of all unmarked sites, and the prime on the sum denotes the exclusion of $i=w$.
The linear part of the Hamiltonian is
therefore a two-level operator in $|w\rangle$ and $|\alpha\rangle$. The 
equation of motion for any site $k\neq w$ takes the form 
$\imath\psi_k=-\gamma\sum_m\psi_m+g|\psi_k|^2\psi_k$, where 
$\psi_k\equiv\langle k|\psi\rangle$. Together with the fact that the initial 
amplitudes are the same on all sites, this means that the amplitudes on all
unmarked sites are identical for all times. The nonlinear term can therefore be
written
\begin{equation}
H_{\rm NL}=g|\psi_w|^2|w\rangle\langle w|+g|\psi_v|^2{\sum_i}'|i\rangle
\langle i|,
\label{eq:HNL}
\end{equation}
where $v$ denotes any vertex such that $v\neq w$. Because $\psi_{\alpha}
=\sqrt{N-1}\psi_v$, the nonlinear term becomes
\begin{equation}
H_{\rm NL}=g|\psi_w|^2|w\rangle\langle w|+\frac{g|\psi_{\alpha}|^2}{N-1}
{\sum_i}'|i\rangle\langle i|.
\end{equation}

The equations of motion can be found using
\begin{equation}
\langle w|\imath\frac{\partial}{\partial t}|\psi\rangle
=\langle w|H|\psi\rangle;\quad
\langle \alpha|\imath\frac{\partial}{\partial t}|\psi\rangle
=\langle \alpha|H|\psi\rangle.
\end{equation}
After straightforward algebra, one obtains
\begin{subequations}
\begin{equation}
\imath \frac{\partial}{\partial t}\psi_{\alpha}= -\gamma (N-1)\psi_{\alpha}
-\gamma\sqrt{N-1}\psi_{w}+\frac{g}{N-1}|\psi_{\alpha}|^2\psi_{\alpha};
\end{equation}
\begin{equation}
\imath \frac{\partial}{\partial t}\psi_{w}= -\gamma\sqrt{N-1}\psi_{\alpha}
-(1+\gamma)\psi_{w}+g|\psi_{w}|^2\psi_{w},
\end{equation}
\label{dyn2}
\end{subequations}
with the initial state
\begin{equation}
\left(\begin{array}{c}
\psi_{\alpha}(0)\\
\psi_w(0)\\
\end{array}\right)=\frac{1}{\sqrt{N}}\left(\begin{array}{c}
\sqrt{N-1}\\
1\\
\end{array}\right).
\label{psinit}
\end{equation}
As hoped, the equations of motion for the complete graph have now been reduced
to a two-level operator. For a large search space ($N\gg1$), Eqs.~(\ref{dyn2}) 
become
\begin{subequations}
\begin{equation}
\imath \frac{\partial}{\partial t}\psi_{\alpha}\approx -\gamma N\psi_{\alpha}
-\gamma\sqrt{N}\psi_{w}+\frac{g}{N}|\psi_{\alpha}|^2\psi_{\alpha};
\end{equation}
\begin{equation}
\imath \frac{\partial}{\partial t}\psi_{w}\approx -\gamma\sqrt{N}\psi_{\alpha}
-(1+\gamma)\psi_{w}+g|\psi_{w}|^2\psi_{w},
\end{equation}
\label{dyn3}
\end{subequations}
with the initial state 
\begin{equation}
\left(\begin{array}{c}
\psi_{\alpha}(0)\\
\psi_w(0)\\
\end{array}\right)\approx
\left(\begin{array}{c}
1\\
0\\
\end{array}\right).
\end{equation} 

Since $\psi_{\alpha}$ and $\psi_w$ are complex variables, 
they can be represented as
\begin{equation}
\psi_{\alpha}\equiv \sqrt{N_{\alpha}} e^{\imath\theta_{\alpha}};\quad
\psi_w\equiv \sqrt{N_w} e^{\imath\theta_w},
\end{equation}
where $N_{\alpha}$ and $N_{w}$ are the populations of bosons in the states 
$|\alpha\rangle$ and $|w\rangle$, respectively. Eqs.~(\ref{dyn3}) then 
correspond to four coupled nonlinear differential equations. To reduce these to two
coupled equations, one can define new variables
\begin{subequations}
\begin{equation}
\eta\equiv |\psi_{w}|^2- |\psi_{\alpha}|^2= N_w- N_{\alpha};
\end{equation}
\begin{equation}
\phi\equiv \theta_w- \theta_{\alpha},
\end{equation}
\end{subequations}
Eqs.~(\ref{dyn3}) then become
\begin{subequations}
\begin{equation}
\dot{\eta}=2 \gamma \sqrt{N} \sqrt{1-\eta^2}\sin(\phi);
\end{equation}
\begin{equation}
\dot{\phi}=\delta -\frac{g}{2}\eta-2 \gamma \sqrt{N}\frac{\eta}{\sqrt{1-\eta^2}}\cos(\phi),
\end{equation}
\label{dyn4}
\end{subequations}
where $\delta$ is
\begin{equation}
\delta\equiv 1- N\gamma- \frac{g}{2}.
\end{equation}
Eqs.~($\ref{dyn4}$) are almost identical to those describing the Josephson 
dynamics of two weakly coupled Bose-Einstein condensates; see for example
Eq.~(2.6) in Ref.~\cite{Raghavan1999}. Note that taking the large-$N$ limit
is not necessary; the general-$N$ case is recovered simply by replacing
$N\to N-1$. The initial conditions for these variables are
\begin{equation}
\big(\eta(0),\phi(0)\big)=\left(-\frac{N-2}{N},0\right)=\left(-1+\frac{2}{N},0
\right).
\label{dyn4init}
\end{equation}
 
\subsection{Complete search: $\delta=0$}
\label{complete}

The complete search corresponds to the evolution of the state from the initial
state~(\ref{psinit}) where $\{|\psi_{\alpha}(0)|^2,|\psi_w(0)|^2\}
=\left\{1-\frac{1}{N},\frac{1}{N}\right\}$,
corresponding to the maximum probability on the superposition of all unmarked
sites, to the state with maximum probability on the marked site, i.e.\
$\{|\psi_{\alpha}|^2,|\psi_w|^2\}=\left\{\frac{1}{N},1-\frac{1}{N}\right\}$.
In other words, the search is complete when $|\eta(t_s)|=|\eta(0)|$. To 
determine if this is possible, it is convenient to interpret Eqs.~(\ref{dyn4}) 
as Hamilton's equations of motion
\begin{equation}
\dot{\eta}=-\frac{\partial H_C}{\partial \phi};\quad
\dot{\phi}=\frac{\partial H_C}{\partial \eta}
\label{Hameqs}
\end{equation}
for some classical Hamiltonian $H_C$. It is straightforward to verify that a
Hamiltonian satisfying both Eqs.~(\ref{Hameqs}) and $(\ref{dyn4})$ is
\begin{equation}
H_C=\delta\eta-\frac{g}{4}\eta^2+2\gamma \sqrt{N}\sqrt{1-\eta^2}\cos(\phi).
\label{HC}
\end{equation}
This has the same form of classical Josephson Hamiltonian~\cite{CJH}. Since 
$\frac{\partial H}{\partial t}=0$, this Hamiltonian is a constant of motion. 
The value of $H_C$ for a trajectory starting at the initial point 
$\{\eta(0),\phi(0)\}=\left\{-1+\frac{2}{N},0\right\}$ must therefore be the 
same for the desired output
$\{\eta(t_s),\phi(t_s)\}=\left\{1-\frac{2}{N},0\right\}$. This is only 
possible if $2\delta\left(1-\frac{2}{N}\right)=0$ or $\delta=0$, which 
corresponds to the homogeneous case. Setting $\delta=0$ for a complete search 
specifies a critical value for $\gamma$:
\begin{equation}
\gamma^*=\frac{2-g}{2N}
\end{equation}
Note that the hopping coefficient must be positive, which requires $g<2$. 

\begin{figure}[t]
\epsfig{file=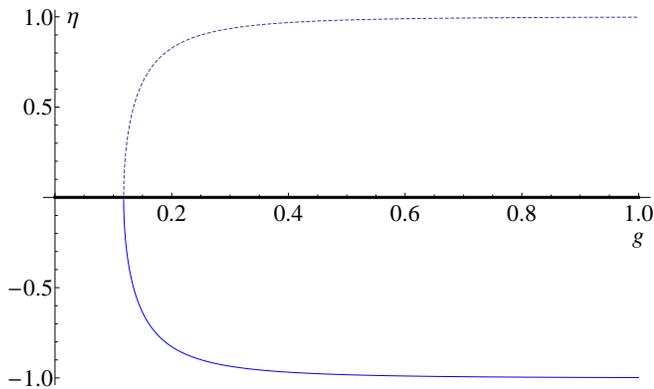,width=\columnwidth}
 \caption{The second set of fixed points as a function of $g$ for $N= 1024$, the dashed line is $\eta_{+}$ and solid line is $\eta_{-}$ and the thick black line is $\eta_{0}$. }
 \label{bifurcation}
\end{figure}

\subsubsection{Fixed points}

The fixed points are obtained by setting Eqs.~(\ref{dyn4}) to zero. These are
\begin{equation}
\phi_1=2m\pi,\;m\in\mathbb{Z};\quad\eta_0=0;
\end{equation}
and
\begin{equation}
\phi_2=(2m+1)\pi;\quad\left\{\begin{array}{c}
\eta_+=+\sqrt{1-\frac{4(g-2)^2}{g^2N}}; \\
\eta_0=0;\hfill\hphantom{a} \\
\eta_-=-\sqrt{1-\frac{4(g-2)^2}{g^2N}}.
\end{array}\right.
\end{equation}
As shown in Fig.~\ref{bifurcation}, $\eta_+$ and $\eta_-$ approach zero as $g$ 
decreases and they both vanish at $g=g^*$, where
\begin{equation}
g^*=\frac{4}{2+\sqrt{N}}.
\end{equation}

\begin{figure}
\epsfig{file=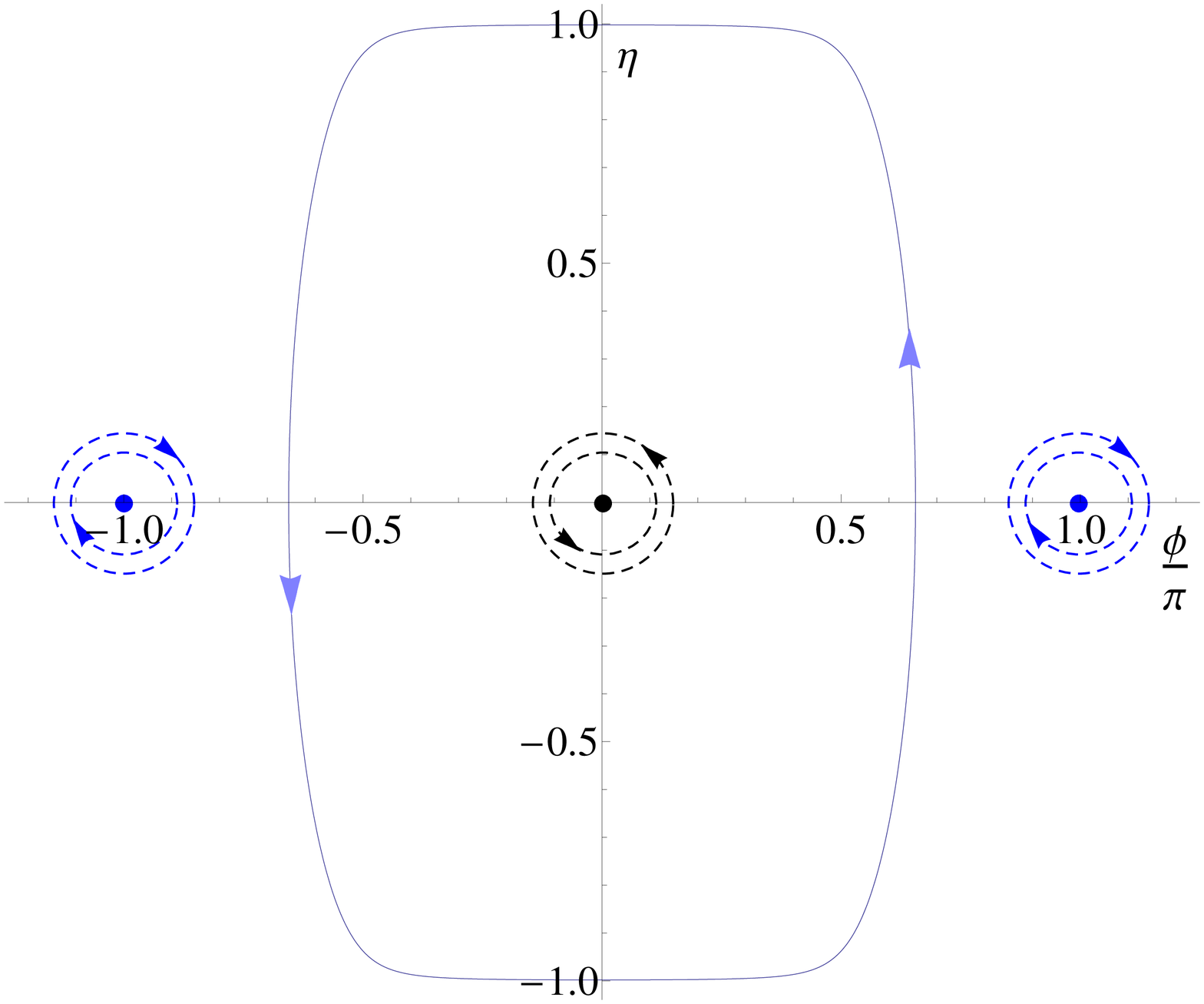,width=\columnwidth}
\caption{Trajectory in phase space for $N=1024$, $g=g^*$, 
$\gamma=\gamma^*=\frac{2-g^*}{2N}$. The black and dark blue dots are 
marginally stable fixed points $(\eta,\phi)=(0,0)$ and $(0,\pm\pi)$,
respectively. Black and dark blue dashed circles are the orbits around these
fixed points. The light blue line corresponds to the trajectory taken for a
complete search.}
\label{orbits}
\end{figure}

To make further progress, one must identify the nature of the fixed points; 
a brief review of these concepts is given in the Appendix.
From Eqs.~(\ref{dyn4}) with $\delta=0$, the functions appearing in the 
Jacobian~(\ref{jacobian}) are 
\begin{subequations}
\begin{equation}
a(\eta,\phi)=2\gamma^* \sqrt{N}\sqrt{1-\eta^2}\sin(\phi);
\end{equation}
\begin{equation}
b(\eta,\phi)=-\frac{g}{2}\eta-2\gamma^*\sqrt{N}\frac{\eta}{\sqrt{1-\eta^2}}\cos(\phi).
\end{equation}
\end{subequations}
The Jacobian is therefore
\begin{equation}
J=\frac{2\gamma^*\sqrt{N}}{\sqrt{1-\eta^2}}\left(\begin{array}{cc}
      -\eta\sin(\phi) & (1-\eta^2)\cos(\phi) \\
        -\frac{g \sqrt{1-\eta^2}}{4\gamma^*\sqrt{N}}
				-\frac{\cos(\phi)}{1-\eta^2} & \eta \sin(\phi) \\
      \end{array}\right).
\end{equation}
For the first set of fixed points $(\eta,\phi)=(0,2m\pi)$, the Jacobian matrix 
becomes
\begin{equation}
J_{(0 , 2m\pi)}= \left(\begin{array}{cc}
      0 & \frac{2-g}{\sqrt{N}} \\
      -\frac{g}{2}-\frac{2-g}{\sqrt{N}}   &0 \\
      \end{array}\right), 
\end{equation}
where the condition $\gamma^*=(2-g)/2N$ has been applied. The eigenvalues are
\begin{equation}
\lambda_{\pm}=\pm i\sqrt{\frac{(2-g)[4+g(\sqrt{N}-2)]}{2N}}.
\end{equation}
Because $N\gg 1$ and $\gamma^*>0$, the eigenvalues are strictly imaginary. The
fixed points are therefore marginally stable, or centers. The orbit in the 
vicinity of the fixed point at $(\eta,\phi)=(0,0)$ is counterclockwise, as
shown in black in Fig.~\ref{orbits}.

\begin{figure}[t]
\epsfig{file=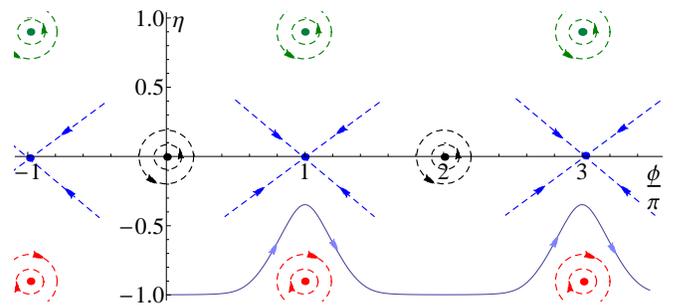,width=\columnwidth}
\caption{Trajectory in phase space with $N=1024$, $g=2g^*$ and 
$\gamma=\gamma^*=(2-g^*)/2N$ (light blue line). The red dots are $\eta_{-}$, 
the green ones are $\eta_{+}$, and the black ones are $\eta_{0}$. The dark 
blue vectors are the eigenvectors of the Jacobian matrix at $(0,(2m+1)\pi)$.}
\label{extended} 
\end{figure}

The second set of fixed points correspond to 
$(\eta,\phi)=(\{\eta_{\pm},0\},(2m+1)\pi)$. Consider first the case
$(\eta,\phi)=(0,(2m+1)\pi)$. The eigenvalues of the Jacobian matrix are then
\begin{equation}
\lambda_{\pm}=\pm\sqrt{\frac{(2-g)[-4+g(\sqrt{N}+2)]}{2N}}.
\end{equation}
If $g>\frac{4}{2+\sqrt{N}}$ these eigenvalues are both real. One is 
negative and the other positive, yielding an unstable saddle fixed point. The
trajectories in the vicinity of these fixed points are depicted as dark blue 
vectors in Fig.~\ref{extended}. If $g<\frac{4}{2+\sqrt{N}}$ then both 
eigenvalues are imaginary, yielding a marginally stable fixed point or center. 
The orbits in the vicinity of the fixed points $(\eta,\phi)=(0,\pm\pi)$ are 
clockwise, as shown in black in Fig.~\ref{orbits}.

Consider next the fixed points $(\eta,\phi)=(\eta_{\pm},(2m+1)\pi)$. For both
cases, the eigenvalues of the Jacobian matrix are
\begin{equation}
\lambda_{\pm}=\pm\frac{1}{2}\sqrt{\frac{4(2-g)^2}{N}-g^2}.
\end{equation}
For $g<\frac{4}{2+\sqrt{N}}$ the solutions $\eta_{\pm}$ do not exist. For 
$g>\frac{4}{2+\sqrt{N}}$, both eigenvalues become imaginary, yielding a
marginally stable fixed point or center. The only difference between the two 
cases is that trajectories near the $\eta_+$ flow counterclockwise, opposite 
to the clockwise direction for those near $\eta_-$; these are shown as green
and red orbits in Fig.~\ref{extended}.

Fig.~\ref{orbits} clearly shows that all of the fixed points in the 
$g\lesssim g^*$ regime are marginally stable or centers. Hence in this regime 
the trajectory starts from the initial point 
$\big(\eta(0),\phi(0)\big)=\left(-1+\frac{2}{N},0\right)$, rotates around the 
origin, and reaches the final point of the search 
$\big(\eta(t_s),\phi(t_s)\big)=\left(1-\frac{2}{N},0\right)$. This behavior is 
depicted as the light blue line in Fig.~\ref{orbits}. Thus, a complete search 
is attainable in this regime.

In the other regime $g>g^*$ there is a saddle fixed point 
$(\eta,\phi)=(\eta_-,\pi)$ near the initial point $(-1,0)$. The trajectory will 
continue along the positive $\phi$ and $\eta$ will remain close to $\eta_-$,
so that it would never reach $\eta=+1$. As shown in Fig.~\ref{extended}, it is 
not likely to have a complete search in this regime. In principle, one might
still have a complete search for $g\gtrsim g^*$ because the linearization 
procedure is strictly valid only right at the fixed points, but the range is
not likely to be extensive. In any case, for any $g\in [0,g^*]$ a complete 
search is attainable. The value $g=g^*$ will be chosen for the remainder of 
the calculations.

\subsubsection{Search time}

Now that the nonlinear quantum search on the complete graph has been shown to
be successful for a range of interaction strengths $g$, it is important to
determine the scaling of the search time $t_s$ with the number of sites for 
large $N$. As shown in Fig.~\ref{orbits}, the trajectory closely resembles a 
rectangle. Consider the dynamical equations~(\ref{dyn4}) with $\delta=0$ and 
the large-$N$ values of $g=g^*\approx 4/\sqrt{N}$ and 
$\gamma=\gamma^*=(2-g)/2N\approx 1/N$:
\begin{subequations}
\begin{equation}
\dot{\eta}=\frac{2}{\sqrt{N}}\sqrt{1-\eta^2}\sin(\phi);
\end{equation}
\begin{equation}
\dot{\phi}=-\frac{2}{\sqrt{N}}\eta\left[1+\frac{1}{\sqrt{1-\eta^2}}\cos(\phi)
\right],
\end{equation}
\label{dyn5}
\end{subequations}
For $N\gg 1$, the right hand sides of Eq.~(\ref{dyn5}) approach zero, and the
trajectories are approximated by $\eta(t)\approx k_1$ and 
$\phi\approx k_2$, with $k_1$ and $k_2$ constants. These can be approximately 
decomposed into the following steps:

\begin{enumerate}

\item[(I)] Constant $\eta$: the initial point, $(\eta,\phi)\approx (-1,0)
\rightarrow (-1,\phi_c)$, where $\phi_c$ is the intersection of the trajectory 
with the $\phi$ axis;

\item[(II)] Constant $\phi$: $(\eta,\phi)\approx (-1,\phi_c)\rightarrow
(1,\phi_c)$;

\item[(III)] Constant $\eta$: $(\eta,\phi)\approx (1,\phi_c) \rightarrow 
(1,0)$.

\end{enumerate}

\begin{figure}[t]
\epsfig{file=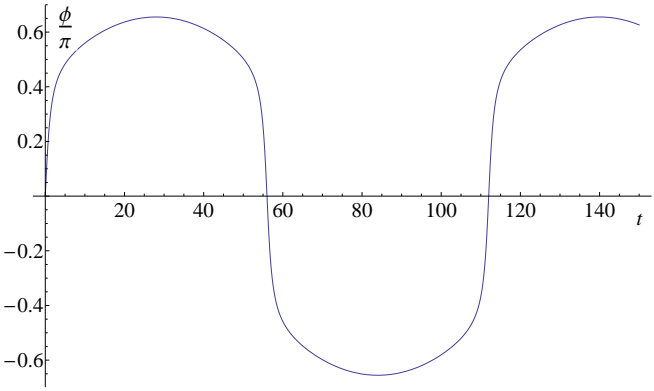,width=\columnwidth}
\caption{The phase $\phi$ as a function of time, when $\delta=0$, 
$g=g^*=\frac{4}{\sqrt{N}}$ and $N=1024$.}
\label{f5} 
\end{figure}

The search time $t_s$ can be written as
\begin{equation}
t_s=\int_{\rm (I)} dt+\int_{\rm (II)} dt+\int_{\rm (III)} dt.
\end{equation}
For the first and third steps $\eta$ is approximately constant, just as $\phi$
is approximately constant for the second step, so
\begin{equation}
t_s\approx\int_0^{\phi_c}\frac{d\phi}{\dot{\phi}}
+\int_{-1}^1 \frac{d\eta}{\dot{\eta}}+\int_{\phi_c}^0 \frac{d\phi}{\dot{\phi}}.
\end{equation} 
The initial 
condition~(\ref{dyn4init}) gives a trajectory $\eta(t)\approx -1$ and 
$\dot{\phi}\to\infty$ so that the first and third integrals make an 
insignificant contribution to $t_s$. As shown in Fig.~\ref{f5}, the 
$\eta\approx k_1$ transition $\phi=0\rightarrow\phi_c$ is much faster than
the $\phi\approx k_2$ transition $\eta=-1\rightarrow 1$ (during which the phase 
hovers in the vicinity of $\phi_c$). This system is an example of a relaxation 
oscillator~\cite{Strogatz1994}. Therefore one can express $t_s$ as
\begin{eqnarray}
t_s&\approx&\int_{-1}^1\frac{d\eta}{\dot{\eta}}\approx
\frac{\sqrt{N}}{2\sin(\phi_c)}\int_{-1}^1\frac{d\eta}{\sqrt{1-\eta^2}}
\nonumber \\
&=&\frac{\pi}{2}\left(\frac{1}{\sin(\phi_c)}\right)\sqrt{N}
\label{ts}
\end{eqnarray}
for large $N$. Besides the factor of $1/\sin(\phi_c)$, this is the usual
expression for the spatial search time.

Setting $\delta=0$, $g=g^*$, and $\gamma=\gamma^*$, the classical Hamiltonian
$H_C$ in Eq.~(\ref{HC}) takes the form:
\begin{equation}
H_C=\frac{2\sqrt{1-\eta^2}\cos(\phi)-\eta^2}{2+\sqrt{N}}.
\label{HC2}
\end{equation} 
For the initial condition $\eta(0)=-1+\frac{2}{N}\approx -1$ for $N\gg 1$, the
classical Hamiltonian becomes $H_C\approx -\frac{1}{\sqrt{N}}$ and this value 
is preserved during the evolution. When the trajectory depicted in 
Fig~\ref{orbits} crosses the $\phi$
axis at the point $(\eta,\phi)=(0,\phi_c)$, the Hamiltonian is approximately
$H_C\approx 2\cos(\phi_c)/\sqrt{N}$. The value of the phase at this point is 
therefore $\phi_c=\cos^{-1}(-1/2)=2\pi/3$. This is consistent with the
time-evolution of the phase shown in Fig.~\ref{f5}. The time for the nonlinear 
search in the large-$N$ limit, Eq.~(\ref{ts}), is therefore
\begin{equation}
t_s=\frac{\pi}{2}\frac{2}{\sqrt{3}}\sqrt{N}=\frac{\pi}{\sqrt{3}}\sqrt{N},
\end{equation}
slower than the linear search by a constant factor of 
$2/\sqrt{3}\approx 1.155$.

It is important to check that the linear search time is recovered in the case
$g\to 0$. In this case one has $\gamma^*=1/N$ (valid for all $N$). The 
trajectories will then cross the $\eta$ axis at the initial condition 
$\big(\eta(0),\phi(0)\big)=\left(-1+\frac{2}{N},0\right)$ when the classical
Hamiltonian~(\ref{HC}) takes the value $H_C\approx\frac{2\sqrt{2}}{N}$,
valid for $N\gg 1$. When the trajectory crosses the $\phi$ axis at 
$(\eta,\phi)=(0,\phi_c)$, the classical Hamiltonian becomes
$H_C=\frac{2}{\sqrt{N}}\cos(\phi_c)$. Because the Hamiltonian is a constant of 
the motion, one obtains 
$\phi_c\approx\cos^{-1}\left(\sqrt{\frac{2}{N}}\right)\approx\pi/2$. The search 
time for the linear problem, Eq.~(\ref{ts}), is then $t_s=(\pi/2)\sqrt{N}$, 
consistent with expectations.

\subsubsection{Errors}

In the foregoing analysis, it has been assumed that the initial amplitudes on 
all sites are always identical, as are the amplitudes to hop from site to site.
The derivation of the nonlinear Hamiltonian~(\ref{eq:HNL}) is only valid under 
these condition. Relaxing these assumptions prevents the reduction of the 
$N$-vertex system to a two-dimensional problem. Instead, one must solve $N$ 
simultaneous coupled nonlinear differential equations of the form
\begin{equation}
\imath\frac{\partial}{\partial t}\psi_j=-\sum_{k=1}^N\gamma_{jk}\psi_k
-\psi_w\delta_{jw}+g|\psi_j|^2\psi_j,
\label{eq:GPnoise}
\end{equation}
where $j=1,\ldots,N$ including the marked site $j=w$. 

\begin{figure}[t]
\epsfig{file=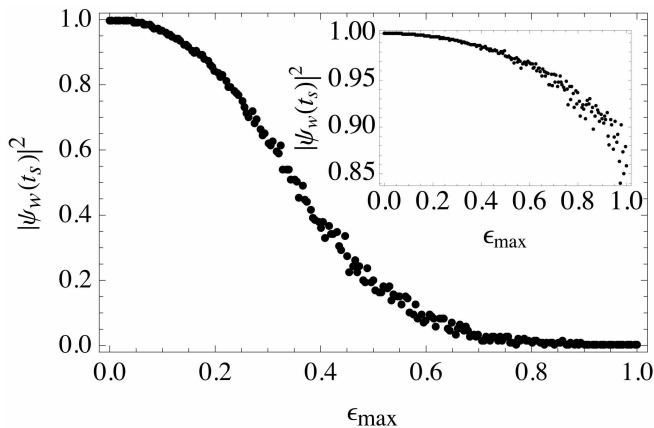,width=\columnwidth}
\caption{Probability of obtaining the marked site $|\psi_w(t_s)|^2$ at the 
search time $t_s=\pi\sqrt{N/3}$ as a function of the maximum error 
$\epsilon_{\rm max}$ in the initial state~(\ref{eq:GPnoiseinit}), for 
$g=4/\sqrt{N}$, $\gamma_{ij}=\gamma^*=(2-g)/2N$, and $N=600$. The inset shows 
the success probability under the same conditions, but assuming that the 
initial state has a constant phase.}
\label{fig:noise}
\end{figure}

Consider first the possibility that the initial state is not the uniform
superposition of all sites, but instead some arbitrary input. Let the initial 
state be
\begin{equation}
|\psi(0)\rangle=\frac{1}{\sqrt{N+\sum_j\epsilon_j}}\sum_{j=1}^N
\sqrt{1+\epsilon_j}e^{\imath \pi \epsilon_j}|j\rangle,
\label{eq:GPnoiseinit}
\end{equation}
where the uniform amplitudes of Eq.~(\ref{init}) on each site have now been 
deformed by a random real number $|\epsilon_j|\leq\epsilon_{\rm max}$ as well
as phases in the range $\{-\epsilon_{\rm max}\pi,\epsilon_{\rm max}\pi\}$.
Eqs.~(\ref{eq:GPnoise}) were solved numerically in Mathematica with 
$g=g^*=4/\sqrt{N}$, $\gamma_{ij}=\gamma^*=(2-g)/2N$, and $t=t_s=\pi\sqrt{N/3}$,
and the resulting 
probability of finding the marked site $|\psi_w(t_s)|^2$ is plotted in 
Fig.~\ref{fig:noise} for the particular case $N=600$. While the data show some
fluctuations due to the randomization, the results clearly indicate that the
marked site can be found with probability exceeding 90\% for errors
$\epsilon_{\rm max}\lesssim 0.15$. Surprisingly, if the initial state is
assumed to possess phase coherence (i.e.\ no randomized phases are included)
then the marked site can be obtained with probability exceeding 90\% for 
$\epsilon_{\rm max}\lesssim 0.8$, as shown in the inset of 
Fig.~\ref{fig:noise}. Note that the $\epsilon_{\rm max}=1.0$ case corresponds
to a completely random (but constant phase) initial state. These results 
indicate that the nonlinear quantum search is robust against initialization 
noise.

\begin{figure}[t]
\epsfig{file=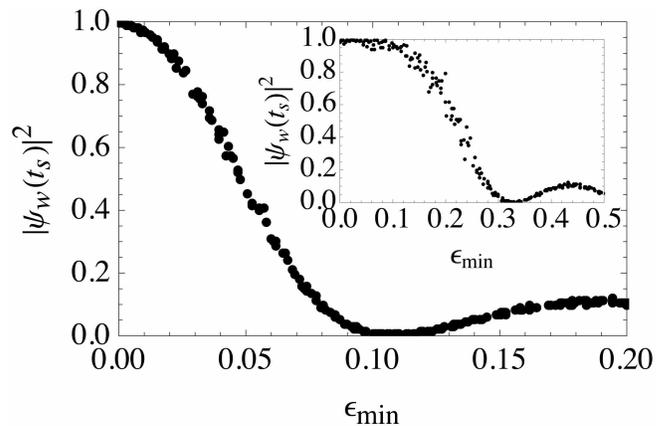,width=\columnwidth}
\caption{Probability of obtaining the marked site $|\psi_w(t_s)|^2$ at the 
search time $t_s=\pi\sqrt{N/3}$ as a function of the number of zero-weight
edges $\epsilon_{\rm min}$, assuming a uniform initial state, $g=4/\sqrt{N}$, 
$\gamma_{ij}=\gamma^*=(2-g)/2N$ for non-zero edges, and $N=300$. The inset 
shows the success probability under the same conditions, but assuming that the 
non-zero edges now have random amplitudes 
$\gamma_{ij}>2\gamma^*\epsilon_{\rm min}$.}
\label{fig:gamnoise}
\end{figure}

Consider second the possibility that the hopping amplitudes $\gamma_{ij}$ can 
vary while the initial state is assumed to be uniform. The simplest case is to
consider the effect of randomly deleting edges, i.e.\ to suppose that 
$\gamma_{ij}=\gamma^*$ for some fraction $1-\epsilon_{\rm min}$ of the edges, 
and is zero otherwise. Given a random variable $\epsilon_{ij}\in\{0,1\}$ then 
one can define
\begin{equation}
\gamma_{ij}=
\begin{cases}
\gamma^* & \epsilon_{ij}>\epsilon_{\rm min} \cr
0 & {\rm otherwise},\cr
\end{cases}
\end{equation}
which produces an Erd\" os-R\' enyi random graph~\cite{Erdos1959} with 
approximately $\left(1-\epsilon_{\rm min}\right)N(N-1)/2$ edges. 
Fig.~\ref{fig:gamnoise} depicts the results for $g=g^*=4/\sqrt{N}$ and $N=300$,
assuming a uniform initial condition. The data show that the quantum search 
success probability drops precipitously as a function of the fraction of zero 
edges $\epsilon_{\rm min}$; to ensure 90\% probability or better on the marked 
site requires $\epsilon_{\rm min}\lesssim 0.02$. The output probability does 
not decrease monotonically with $\epsilon_{\rm min}$, but in fact increases 
again slightly for as the number of zero-weight edges increases beyond 
approximately 10\%. Similar observations of enhanced quantum search with 
increased connectivity have been reported in the context of the linear
discrete-time quantum walk~\cite{Lovett2011}.

Another model for including error in the structure of the graph is to assume
that the values of the hopping amplitudes are not constant. Consider for
concreteness the case $\gamma_{ij}=(1+\epsilon_{ij})\gamma^*$ with the random
variable $|\epsilon_{ij}|\leq 1$. The numerics yield the surprising result that
the probability of finding the marked site $|\psi_w(t_s)|^2\to 1$ as 
$N\to\infty$ (not shown). For large graphs, the success of the algorithm is 
therefore not affected by
the randomization of the hopping amplitudes, as long as their average is
$\gamma^*$ and the initial state is assumed to be uniform. To make contact with 
the results on Erd\" os-R\' enyi graphs, suppose that in addition to the 
randomization of the hopping amplitudes one adds the supplementary condition 
that $\gamma_{ij}=0$ if $1+\epsilon_{ij}<2\epsilon_{\rm min}$ (recall that 
$0\leq\gamma_{ij} \leq 2\gamma^*$). In this case $\epsilon_{\rm min}$ again
reflects the fraction of edges that have zero weight, while the remaining edges 
have random amplitudes $\gamma_{ij}>2\gamma^*\epsilon_{\rm min}$. The results, 
depicted in the inset of Fig.~\ref{fig:gamnoise}, indicate that randomizing the 
hopping amplitudes in
this way in fact improves the algorithm's success probability under edge
deletion relative to the unweighted case. Under these conditions, an output 
probability exceeding 90\% can be obtained with approximately 12\% of the 
complete graph's edges deleted, compared with only approximately 2\% in the 
unweighted case.

\subsection{Incomplete search: $\delta\neq0$}

As discussed at the 
beginning of Sec.~\ref{complete}, a complete search is equivalent to finding a
trajectory that makes the transition $\eta=-1+\frac{2}{N}\to 1-\frac{2}{N}$ 
possible. Because $H_C$ is a constant of the motion, however, a complete
search requires $\delta=0$. That said, it is conceivable that setting 
$\delta\neq 0$ could yield a time $t_s$ 
where the relative occupation of the marked site would instead be 
$\eta\lesssim 1-\frac{2}{N}$. Consider again the fixed
points of Eqs.~(\ref{dyn4}). Evidently $\phi=m\pi$ ensures that the right hand 
side of Eq.~(\ref{dyn4}a) is zero, but $\eta=0$ no longer accomplishes this for
the right hand side of Eq.~(\ref{dyn4}b) because $\delta=1-N\gamma-g/2\neq 0$ 
by assumption. Now that the search can not be complete, it is not necessary to 
keep the exact expressions for finite $N$, and one can work entirely in the 
limit $N\gg 1$. If one again sets $\gamma=\gamma^*=1/N$, then $\delta=-g/2$. 
Choosing $g=\alpha/N^{\beta}$, where $\alpha$ and $\beta$ are both positive 
real numbers, gives $\delta\sim N^{-\beta}$ which vanishes for large $N$. 
Eqs.~(\ref{dyn4}) then become
\begin{subequations}
\begin{equation}
\dot{\eta}=\frac{2\sqrt{1-\eta^2}}{\sqrt{N}}\sin(\phi);
\end{equation}
\begin{equation}
\dot{\phi}=-\frac{\alpha(1+\eta)}{2N^{\beta}}
-\frac{2\eta}{\sqrt{N}\sqrt{1-\eta^2}}\cos(\phi).
\end{equation}
\label{dyn6}
\end{subequations}

To determine the maximum value reached by $\eta$, consider the 
classical Hamiltonian~(\ref{HC}) which now has the form
\begin{equation}
H_C=-\frac{\alpha\eta(2+\eta)}{4N^{\beta}}+\frac{2\sqrt{1-\eta^2}}{\sqrt{N}}
\cos(\phi).
\label{HCinc}
\end{equation}
At the initial conditon $(\eta,\phi)\approx(-1,0)$, the classical Hamiltonian 
is $H_C(0)=\alpha/4N^{\beta}$ which is a constant of the motion. Note that for 
the incomplete search, the classical Hamiltonian is now positive. The 
accessible 
values of $(\eta,\phi)$ are found by setting $H_C=H_C(0)$. The $N$-dependence
disappears if $\beta=1/2$ (i.e.\ $g=\alpha/\sqrt{N}$), and the only two real 
solutions correspond to $\eta_1=-1$ and
\begin{equation}
\eta_2=-1+\frac{4(3x)^{1/3}}{3\alpha^2}-\frac{16\cos^2(\phi)}{(3x)^{1/3}},
\label{hinc}
\end{equation}
where
\begin{eqnarray}
x&=&\alpha^3\cos^2(\phi)\sqrt{3}\sqrt{27\alpha^2+64\cos^2(\phi)}
\nonumber \\
&+&9\alpha^4\cos^2(\phi).
\end{eqnarray}
Though the solution~(\ref{hinc}) is a bit unwieldy, a few observations can be 
immediately made. For small and large $\alpha$, one obtains respectively
\begin{eqnarray}
\eta_2(\alpha\ll 1)&\approx& 1-\frac{\alpha^2}{8}\sec^2(\phi);
\label{alphasmall} \\
\eta_2(\alpha\gg 1)&\approx& -1+4\left(\frac{2\cos^2(\phi)}{\alpha^2}
\right)^{1/3}.
\label{alphabig}
\end{eqnarray}
Note that only even powers of $\alpha$ appear in these expansions, indicating
that the dynamics is unaffected by the sign of the nonlinearity. 
Eq.~(\ref{alphasmall}) shows that a complete search is only possible for the
non-interacting case $\alpha=0$, which is consistent with $\delta=0$. For 
any finite interaction strength in this $\delta\neq 0$ regime, a complete
search is not possible. The relative fraction on the marked site decreases 
with $\alpha$, and Eq.~(\ref{alphabig}) reveals that it is asymptotically zero 
for very large nonlinearities (though one still requires $g<2$). Physically,
the large interaction between atoms favors the initial state which is the 
superposition of occupying all sites of the graph. This is the dynamical
self-trapping which has been noted previously for interacting Bose-Einstein
condensates~\cite{Anker2005,Alexander2006}.

Interestingly, $\eta_2^{\rm (max)}$ is independent of the size of the search 
problem (keep in mind however that $g=\alpha/\sqrt{N}$ so that the strength of 
the nonlinearity decreases with $N$). Note also that the maximum probability is
reached for $\phi=0$. For example, $\eta_2^{\rm (max)}=0.99$ requires
$\alpha\approx 0.28$. The case $\eta_2=0$ can be obtained directly from 
Eq.~(\ref{hinc}), and one obtains $\alpha=\pm 8\cos(\phi)$. At this value of 
$\alpha$, the probability of occupying the marked site exactly $1/2$; for any 
smaller $\alpha$ it is larger.

It remains to calculate the time for the incomplete search $t_s$. As in the 
$\delta=0$ case, the right hand sides of Eqs.~(\ref{dyn6}) approach zero, 
which means that the trajectories are approximated by constant lines. Again, 
the $\phi$ evolution is much faster than the $\eta$ evolution for the
$\eta\approx\eta_1=-1$ trajectory. The $\eta\approx\eta_2$ trajectory is not
going to be as fast because the $\sqrt{1-\eta_2}$ term in Eq.~(\ref{dyn6}b) is
no longer almost zero. The search time is then approximately
\begin{eqnarray}
t_s&\approx&\int_{-1}^1 \frac{d\eta}{\dot{\eta}}
+\int_{\phi_c}^0 \frac{d\phi}{\dot{\phi}}\nonumber \\
&=&\frac{\pi\sqrt{N}}{2\sin(\phi_c)}+\int_{\phi_c}^0\frac{d\phi\sqrt{N}}
{-\frac{\alpha(1+\eta)}{2}-\frac{2\eta}{\sqrt{1-\eta^2}}\cos(\phi)}.\quad
\label{tsinc}
\end{eqnarray} 
The critical angle
$\phi_c$ is obtained by equating the classical Hamiltonian~(\ref{HCinc}) for 
$(\eta,\phi)=(0,\phi_c)$ with $H_C(0)$; this gives 
$\alpha/4\sqrt{N}=2\cos(\phi_c)/\sqrt{N}$ or 
\begin{equation}
\phi_c=\cos^{-1}\left(\frac{\alpha}{8}\right)\approx\frac{\pi}{2}
-\frac{\alpha}{8}
\end{equation}
for small $\alpha$. For the linear case ($\alpha=0$), the critical angle 
coincides with that found in the previous section for the complete search. 
The first term in Eq.~(\ref{tsinc}) is therefore $\pi\sqrt{N}/2\sin(\phi_c)
\approx(\pi\sqrt{N}/2)/\sqrt{1-\alpha^2/64}$ which restricts the strength of
the nonlinearity to $|\alpha|<8$ (recall that for $|\alpha|>8$ the probability
on the marked state is less than one half).

The second term can be simplified by assuming that the value of $\eta$, given
in Eq.~(\ref{hinc}) remains approximately constant over the range of 
integration $\{-\phi_c,0\}$. In fact, plotting $\eta$ for a range of $\alpha$
shows that it varies from 0 to its maximum value, Eq.~(\ref{alphasmall}), but 
this increase occurs only over a very small region in the vicinity of 
$\phi=\phi_c$. For small $\alpha$, the time for the incomplete search, given 
by Eq.~(\ref{tsinc}), can be approximated as
\begin{eqnarray}
t_s&\approx&\frac{\pi}{2}\sqrt{N}\left(1+\frac{\alpha^2}{128}\right)
-\sqrt{N}\int_{-\phi_c}^0\frac{d\phi}{\alpha+4\cos(\phi)/\alpha}\nonumber \\
&\approx&\frac{\pi}{2}\sqrt{N}\left[1+\frac{\alpha}{2\pi}\ln\left(
\frac{\alpha}{16}\right)+\frac{\alpha^2}{128}\right].
\end{eqnarray}
Because the $\alpha$-dependent correction terms are negative in the regime
$0<\alpha<8$, the time for the incomplete search is generally shorter than that
for the complete search by a small factor dependent on the strength of the
nonlinearity.

\section{Conclusions}
\label{concl}

In this work we considered the spatial search algorithm on lattice with the 
topology of the complete graph, under the assumption that the continuous-time 
quantum walk is effected by a zero-temperature Bose-Einstein condensate. In the 
mean-field approximation, the equations of motion become nonlinear and 
correspond to the discrete Gross-Pitaevskii equation. The analytical results,
using methods in nonlinear dynamics and numerical calculations, indicate that a 
complete spatial search remains possible even in the presence of nonlinearity. 

For a successful search,
the nonlinear coupling constant must decrease with the system size as 
$N^{-1/2}$ and the inter-site hopping amplitude decreases as $N^{-1}$, where 
$N$ is the number of sites. The latter condition coincides with the criterion 
for a complete search found for the linear search problem on the complete 
graph~\cite{Childs2004a}. Under these circumstances, the search time is found 
to scale as $t_s\propto\sqrt{N}$, with an overall constant factor that depends 
weakly on the strength of the nonlinearity. The probability of success 
generically decreases with the strength of nonlinearity, but there are 
particular choices of parameters where the success probability can be made
unity. The numerical results further indicate that random errors in the input 
state amplitudes and the hopping amplitudes are not deleterious for the 
performance of the algorithm, but that the inclusion of phase errors in the
input state or edge deletions quickly erode the probability of finding the
marked site. Overall, the quantum search is found to be robust to error under 
a variety of conditions. It would be interesting to explore the influence of 
non-unitary (thermal) noise on the performance of the nonlinear search. In the 
limit of zero nonlinearity, the present results in the absence of error 
completely recover those of Ref.~\cite{Childs2004a}.

The dynamics of the nonlinear system agree closely with those of the linear
quantum walk. This indicates that nonlinearity, as long as its strength is
kept bounded for a given system size, is no impediment to the implementation 
of a quantum spatical search. The results suggest that Bose-Einstein 
condensates consisting of huge numbers of particles can be candidates for the 
implementation of useful quantum algorithms. The hopping amplitude and the 
strength of the nonlinearity need to be adjusted for a given size of the 
search space. In ultracold atomic gases, for example, in principle the former 
could be accomplished by adjusting the depth or spacing of a 
lattice~\cite{Bloch2008}, and the latter through the use of Feshbach 
resonances~\cite{Fattori2008}, though constructing a lattice with the
connectivity of the complete graph it is not currently feasible in these
systems.

It would be 
preferable in practice to conduct the spatial search on a regular lattice, for 
example a square lattice in three or lower dimensions. Unfortunately, 
continuous-time quantum walks based on the ordinary discrete Schr\" odinger 
equation do not provide useful quantum speed-ups on these lattices over the 
classical search time, though discrete-time quantum walks can be constructed
that do~\cite{Ambainis2005}. That said, if the particle dispersion relation is
linear rather than quadratic, the full quantum speed-up is achievable on a 
three-dimensional square lattice~\cite{Childs2004b}. One practical strategy to
achieve this is to artifically induce Dirac fermions by suitably preparing an
optical lattice~\cite{Hou2009}. More directly, the excitations of a 
zero-temperature weakly interacting BEC on a lattice are characterized by a 
linear dispersion relation, which is a hallmark of the underlying 
superfluidity in these systems~\cite{PethickSmith}. It is therefore conceivable
that the nonlinear mean-field dynamics of the BEC on a cubic lattice would 
yield the full quantum speed-up for the continuous-time spatial search problem.
This possibility will be explored in future work.

During the preparation of this manuscript, we became aware of other work that
investigated the same system addressed in the present study~\cite{Meyer2013}.
While their results are consistent with ours when there is overlap, in their
work the strength of the nonlinear coupling is assumed to vary with time. 
Their methods and conclusions are therefore complementary to ours.

\begin{acknowledgments}
The authors are grateful to David Meyer for originally stimulating this line 
of inquiry and for productive conversations, and to J\" orn Davidsen and 
Dennis Salahub for their comments on a preliminary manuscript. This work was 
supported by the Natural Sciences and Engineering Research Council of Canada.
\end{acknowledgments}

\appendix*
\section{Characterizing fixed points}
\label{app}

This discussion follows Ref.~\cite{Strogatz1994}. Consider a general 
two-dimensional nonlinear system of equations:
\begin{equation}
\nonumber \dot{u}=a(u,v);\quad \dot{v}=b(u,v).
\label{coupledeqs}
\end{equation}
A fixed point or equilibrium point is defined by
\begin{equation}
\dot{w}\equiv\left(\begin{array}{cc}
         \dot{u} \\
          \dot{v}\\
         \end{array}\right)\equiv 0.
\end{equation} 
Suppose $(u^*,v^*)$ is a fixed point for this system, satisfying 
$a(u^*,v^*)=b(u^*,v^*)=0$.
Let $u^\prime=u-u^*$ and $v^\prime=v-v^*$.
For small $u^\prime$ and $v^\prime$
\begin{subequations}
\begin{equation}
\dot{u^\prime}\approx a(u^*,v^*)+\frac{\partial a}{\partial u}(u^*,v^*)
u^\prime+\frac{\partial a}{\partial v}(u^*,v^*) v^\prime;
\end{equation}
\begin{equation}
\dot{v^\prime}\approx b(u^*,v^*)+\frac{\partial b}{\partial u}(u^*,v^*)
u^\prime+\frac{\partial b}{\partial v}(u^*,v^*)v^\prime.
\end{equation}
\label{linear1}
\end{subequations}
Since $(u^*,v^*)$ is a fixed point, $a(u^*,v^*)=b(u^*,v^*)=0$. Close to the 
fixed points ($u^\prime\ll 1,v^\prime\ll 1$), Eqs.~(\ref{linear1}) can be 
written as
\begin{equation}
\dot{w}^{\prime}\approx Jw^{\prime},
\label{linear2}
\end{equation}
where 
\begin{equation}
w^{\prime}=\left(\begin{array}{cc}
        u^\prime \\
         v^\prime\\
        \end{array}\right),
\end{equation}
and $J$ is the Jacobian matrix 
\begin{equation}
J= \left(\begin{array}{cc}
       \frac{\partial a}{\partial u} &  \frac{\partial a}{\partial v} \\
        \frac{\partial b}{\partial u} & \frac{\partial b}{\partial v} \\
      \end{array}\right)_{(u^*,v^*)}.
\label{jacobian}
\end{equation}

The general solution of Eq.~(\ref{linear2}) when $J$ is non-degenerate and 
invertible is
\begin{equation}
w(t)=c_1z_1e^{\lambda_1t}+c_2z_2e^{\lambda_2t},
\end{equation}  
where $\lambda_i$ and $z_i$ are the eigenvalues and eigenvectors of the 
Jacobian matrix $J$, respectively; $c_1$, $c_2$ are constants which are 
determined by the initial conditions. There are four possibilities for the 
stability of the fixed points:
\begin{enumerate}

\item $\lambda_1$ and $\lambda_2$ are both real:

\begin{enumerate}
\item {\bf Stable fixed point}: If $\lambda_1<0$ and $\lambda_2<0$, then 
$w'\rightarrow 0$ as $t\rightarrow\infty$.

\item {\bf Unstable fixed point}: If $\lambda_1>0$ and $\lambda_2>0$, then 
$w'\rightarrow\infty$ as $t\rightarrow\infty$.

\item {\bf Unstable saddle point}: If $\lambda_1<0<\lambda_2$, then if $w'(0)$
is a multiple of $z_1$ then $w'\rightarrow 0$ as $t\rightarrow\infty$ (stable
along direction of $z_1$); alternatively if $w'(0)$ is a multiple of $z_2$ 
then $w'\rightarrow\infty$ as $t\rightarrow\infty$ (unstable along direction 
of $z_2$).

\end{enumerate}

\item {\bf Marginally stable fixed point / center}: $\lambda_1$ and $\lambda_2$ 
are both complex. If $\Re(\lambda_i)=0$ then $|w'|\to\mbox{const}$ as 
$t\to\infty$. Trajectories circulate around the fixed point and eventually 
return to the initial point; these are closed orbits. 

\end{enumerate}

The Hartman-Grobman theorem states that the dynamics of the linearized system 
in the vicinity of hyperbolic fixed points, where $\Re(\lambda_i)\neq 0$, will 
be similar to that of the original nonlinear system. If one or both eigenvalues 
don't satisfy this condition, the fixed point is non-hyperbolic and therefore
the dynamics are fragile to the inclusion of nonlinearity. That said, suppose 
that $(u^*,v^*)$ is an isolated fixed point that is marginally stable, and 
there exists a conserved quantity $H_C(u,v)$. If $(u^*,v^*)$ is a local minimum 
of $H_C(u,v)$, then all the trajectories sufficiently close to $(u^*,v^*)$ are 
closed.

\end{document}